\documentclass[finalprint, 5p]{elsarticle}
\usepackage{graphicx}
\usepackage{amsmath}
\usepackage{dcolumn}
\usepackage{makeidx}

\begin{document}

\begin{frontmatter}

\title{Ground state of the one dimensional Gross Pitaevskii equation with a Morse potential}
\author[label1]{Sukla Pal\corref{cor1}}
\ead{sukla@bose.res.in}
\author[label1,label2]{J. K. Bhattacharjee}
%\ead{jkb@hri.res.in}
\address[label1]{Department of Theoretical Physics, S.N.Bose National Centre For Basic Sciences, JD-Block, Sector-III, Salt Lake City, Kolkata-700098, India}
\address[label2]{Harish-Chandra Research Institute, Chhatnag road, Jhunsi, Allahabad-211019, India}
\cortext[cor1]{Corresponding author}
\begin{abstract}
We have studied the ground state of the Gross-Pitaevskii equation (nonlinear Schrodinger equation) for a Morse potential via a variational approach. It is seen that the ground state ceases to be bound when the coupling constant of the nonlinear term reaches a critical value. The disappearence of the ground state resembles a saddle node bifurcation.
\end{abstract}
\end{frontmatter}
\section{Introduction}
\label{1}

The nonlinear Schrodinger equation (NLSE) has been a particularly active field of research in condensed matter physics in the last two decades because of its close relationship to the Gross Pitaevskii equation (GPE)\cite{a,b} which describes the dynamics of the condensate in the process of Bose Einstein Condensation (BEC)\cite{7,11}. In the quasi–1D regime, GPE reduces to the one-dimensional NLSE with an external potential. This regime holds when the transverse dimensions of the condensate are of the order of its healing length and the longitudinal dimension is much longer than its transverse dimensions. In this regime the BEC remains phase coherent and the governing equations are one dimensional. This is in contrast to a truly 1D mean-field theory which requires transverse dimensions of the order of or less than the atomic interaction length. In this paper we will deal with 1D GPE and hence with NLSE with Morse type of external potential in x-direction.

Traditionally the evolution equation of a complex variable $\psi(x,t)$ has been called NLSE when it has the structure $i\hbar\frac{\partial\psi}{\partial t}=-\frac{\hbar^2}{2m}\nabla^2\psi+g|\psi|^2\psi$. It becomes the GPE when it has a potential (trapping potential in the case of BEC) V(x) on the right hand side. The most common trapping potential is the simple harmonic oscillator $V(x)=\frac{1}{2}m\omega^2x^2$ which gives a localised ground state. For $V(x)=0$, one possible set of solutions of the NLSE is $\psi(x,t)=\sqrt{\frac{N}{L}}e^{-\frac{iE_kt}{\hbar}}e^{ikx}$, which is the usual free particle plane wave solution and gives the energies $E_k=\frac{\hbar^2k^2}{2m}+\frac{gN}{L}$. This energy spectrum has a gap which violates the Hugenholtz Pines theorem for the interacting Bose gas (repulsive interaction). This motivated us to look for a potential $V(x)$ which can be tuned from a harmonic oscillator potential to a free particle one and study the evolution of the ground state with the evolution of the potential. The potential concerned is the Morse potential $V(x)=D(e^{-2ax}-2e^{-ax})$; where the constant $D$ carries the dimension of energy and `$a$' carries the dimension of inverse length. If $a\rightarrow 0$, then the potential tends to the constant value $-D$ which is effectively a free particle and for $a\rightarrow\infty$, we have a strongly confining potential. So we can tune the potential by tunning `$a$' and accordingly we will see a transition in the nature of the ground state of the particle.

Many applications of the NLSE to BEC’s or GPE have dealt with the ground-state properties as well as various intriguing features of BEC in harmonic oscillator potential \cite{12}, double well potential \cite {c,c1}, anisotropic potential \cite{15}, optical lattice \cite{16,c2}, etc and has been successfully extended to trapped dipolar BEC \cite{13,14}.  There is also growing interest in the possibility of generating topological excitations of a condensate, which may well be described by excited-state solutions of the NLSE \cite {d}. However, in this paper we will concentrate on the ground state properties of GPE in well known Morse potential. We are already familiar with the existence of localized ground state of GPE in presence of harmonic trap and after switching off the trapping potential the rapid expansion of BEC leading no longer to a localized state \cite{9}. 

We present the results of a variational calculation in sec 2 and end with some observation in sec 3.

\section{Variational Calculation}
The one dimensional GPE describing the dilute bosons with repulsive interaction under a trapping potential $V_{ext}$ is given by
\begin{eqnarray}\label{eq1}
i\hbar\partial_t\psi =-\frac{\hbar^2}{2m}\nabla^2\psi+g|\psi|^2\psi+V_{ext}(x)\psi
\end{eqnarray}
with 
\begin{eqnarray}\label{eq2}
\int_{-\infty}^{\infty}|\psi(x)|^2 dx=N
\end{eqnarray}
 N is the total number of particles trapped. Eq. (\ref{eq1}) can be written as $i\hbar\frac{\partial\psi}{\partial t}=\frac{\delta E[\psi]}{\delta \psi}$. Where $E[\psi]$ is the energy functional:
\begin{eqnarray}\label{eq3}
E[\psi(x)]=\int^{\infty}_{-\infty}[\frac{\hbar^2}{2m}|\nabla\psi|^2+V_{ext}|\psi|^2+\frac{g}{2}|\psi|^4]dx
\end{eqnarray}
We will write Morse potential of sec I in the form:

\begin{eqnarray}\label{eq4}
V(bx)=D(e^{\frac{-2bx}{k}}-2e^{\frac{-bx}{k}})
\end{eqnarray}
Where, `D' is the well depth having the dimension of energy and `b' is an inverse confining length $b=\sqrt{\frac{8mD}{\hbar^2}}$, related to `$a$' by $b=ak$, where, $k$ is a dimensionless number. Hereafter we will keep $D$ fixed and vary the confining length by varying the dimensionless number $k$. We have considered a variable transfer of the form $y=ke^{-ax}$ where $k^2=\frac{8mD}{\hbar^2 a^2}$. We fix $D$ at 1.
\begin{figure}[!htbp]
\begin{center}
\includegraphics[angle=0,scale=0.5]{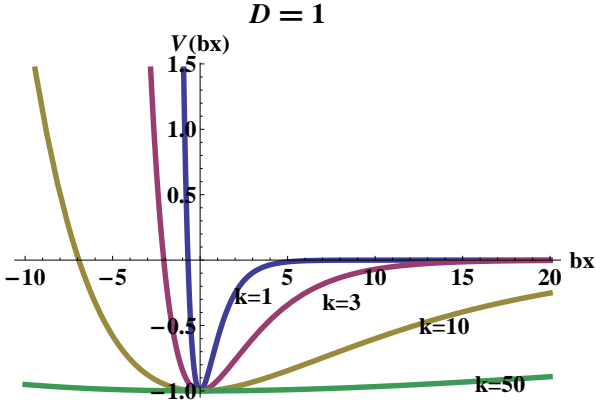}
\caption{Figure above shows the plot of $V(bx)$ vs $bx$ for different values of $k$. This clearly shows that with the change of $k$, the nature of the potential changes.}
\label{fig1}
\end{center}
\end{figure}
For large values of $k$ system becomes essentially free though strictly bounded in the left ($x<0$) at a large distance. As the value of $k$ increases the change in the shape of the potential given in Fig. \ref{fig1} suggests that the system falls under more weaker trap.\\

We try a variational form for the ground state as:
\begin{eqnarray}\label{eq5}
\psi(y)=N_cy^{\alpha}e^{-\beta y}
\end{eqnarray}
After substituting Eq. (\ref{eq5}) into the energy functional $E[\psi]$, Eq. (\ref{eq3}) reduces to 
\begin{eqnarray}\label{eq6}
E[\psi(y)]&=&Na^2\Big(\frac{\alpha}{2}+\frac{\alpha^2}{4\beta^2}+\frac{\alpha}{8\beta^2}-\frac{\alpha k}{2\beta}\Big)\nonumber\\&~&+\frac{mgaN^2}{2^{4\alpha}\hbar^2}\frac{\Gamma(4\alpha)}{\Gamma(2\alpha)^2}
\end{eqnarray}
We consider separately two cases.
\subsection{$g=0$}

Bosons are non interacting in this case and hence Eq. (\ref{eq6}) takes the simple form given below:
\begin{eqnarray}\label{eq7}
E[\psi(y)]_{g=0}= Na^2\Big(\frac{\alpha}{2}+\frac{\alpha^2}{4\beta^2}+\frac{\alpha}{8\beta^2}-\frac{\alpha k}{2\beta}\Big)
\end{eqnarray}
differentiation with respect to the parameters gives following two constraints 
\begin{eqnarray}
1+\frac{\alpha}{\beta^2}+\frac{1}{4\beta^2}=\frac{k}{\beta}\label{eq8}\\
\alpha+\frac{1}{2}=\beta k\label{eq9}
\end{eqnarray}
With the help of Eq. (\ref{eq8}) and Eq. (\ref{eq9}), $\beta$ comes out to be equal to 0.5 and hence independent of $k$. If we substitute this value of $\beta$ in Eq. (\ref{eq8}), we will readily find $\alpha=\frac{k-1}{2}$. Hence the ground state wave function becomes 
\begin{eqnarray}\label{eq10}
\psi_{g=0}(y)|_{\tiny{variation}}=[\frac{aN}{\Gamma (k-1)}]^{1/2}y^{\frac{k-1}{2}}e^{-y/2}
\end{eqnarray}
with ground state energy
\begin{eqnarray}\label{eq11}
E_{g=0}|_{variation}=-\frac{N\hbar^2 a^2}{2m}\big(\frac{1}{2}-\frac{\sqrt{2mD}}{a\hbar})^2
\end{eqnarray}
Eq. (\ref{eq10}) and Eq. (\ref{eq11}) with $N=1$ gives the exact ground state energy as obtained by solving (by applying Laplace and inverse Laplace transformation) Schrodinger Equation for a single particle in Morse potential \cite{5}. 
\subsection{$g\ne0$}

In this case bosons have repulsive interaction. Differentiation with respect to parameter gives the following two constraint conditions in this case.
\begin{eqnarray}\label{eq12}
\alpha+\frac{1}{2}&=&\beta k
\end{eqnarray}
\begin{eqnarray}
\begin{aligned}\label{eq13}
1-\frac{1}{4\beta^2}&=& \frac{kg'}{\sqrt{2}}\frac{\Gamma(4\alpha)}{2^{4\alpha}[\Gamma(2\alpha)]^2}\Big[2\Psi(2\alpha)+4ln2\\ &~&-\Psi(4\alpha)\Big]
\end{aligned}
\end{eqnarray}
Where, $g'=\frac{\sqrt{m}gN}{\hbar\sqrt{D}}$ and $g=\frac{2\hbar^2}{ma_s}$ \cite{6}, $a_s$ being the s wave scattering length and $g'$ being the dimensionless quantity. $\Psi$ is the digamma function defined as $\Psi(x)=\frac{d}{dx}ln[\Gamma(x)]$. To solve Eq. (\ref{eq12}) and (\ref{eq13}) we proceed in the following way. At first the expression of $\beta$ from Eq. (\ref{eq12}) is substituted into Eq. (\ref{eq13})
\begin{eqnarray}
1-\frac{k^2}{(2\alpha+1)^2}&=&\frac{kg'}{\sqrt{2}}C(\alpha)\Big[2\Psi(2\alpha)+4ln2-\Psi(4\alpha)\Big]\nonumber\\
\Rightarrow f_1(\alpha)&=&g'f_2(\alpha)\label{eq14}
\end{eqnarray}
Where, $C(\alpha)=\frac{\Gamma(4\alpha)}{2^{4\alpha}[\Gamma(2\alpha)]^2}$, $f_1(\alpha)$=$1-\frac{k^2}{(2\alpha+1)^2}$ and $f_2(\alpha)=\frac{k}{\sqrt{2}}C(\alpha)\Big[2\Psi(2\alpha)+4ln2-\Psi(4\alpha)\Big]$ with energy functional given below:
\begin{eqnarray}
\begin{aligned}\label{eq15}
\frac{E[\psi(y)]}{ND}&=&\frac{4}{k^2}\Big(\frac{\alpha}{2}+\frac{\alpha^2}{4\beta^2}+\frac{\alpha}{8\beta^2}-\frac{\alpha k}{2\beta}\Big)\\&~&+C(\alpha)\frac{kg'}{2\sqrt{2}}
\end{aligned}
\end{eqnarray}

All the energy values obtained from Eq. (\ref{eq15}) will be in scale of $ND$. Now we will plot both $f_1$ and $f_2$ as a function of $\alpha$ for constant values of $k$ and $g'$ and we will search for the point of intersection.
\section{Observations}

Here we consider fixed value of $k$ at first and then we analyse the case with different values of $k$, i.e., by tunning the shape of the trapping potential.
\subsection{ $k$ is fixed at a constant value: $(k=3.0)$}

For $g'$ close to 0, there is a pair of solutions for $\alpha$. Keeping $k$ fixed at 3.0 as $g'$ is increased, there is a critical value of $g'$ $(= g'_c)$ at which there is only one solution of $\alpha$. Beyond $g'_c$ there is no solution for any value of $g'$. Hence the solutions bifurcate at $g=g_c$. 
\begin{figure}[!htbp]
\begin{center}
\includegraphics[angle=0,scale=0.5]{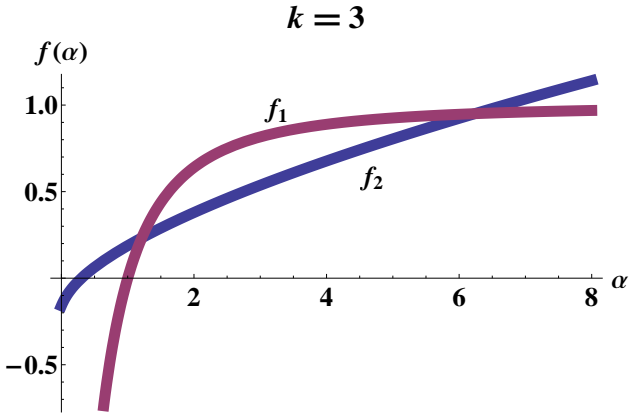}
\caption{In this figure we have plotted $f_1(\alpha)$ and $f_2(\alpha)$ vs. $\alpha$ for $k=3.0$ and $g'=0.1 $. It shows that we have two possible values of $\alpha$ ($\alpha_1=1.2 $and $\alpha_2=6.2 $). This qualitative nature (i.e, two sets of solution for $\alpha$ and $\beta$) of the above figure remains same up to $g'<g'_c$}
\label{fig2}
\end{center}
\end{figure}
From this two values of $\alpha$ ($\alpha_1= 1.2$, $\alpha_2= 6.2$) as shown in Fig. \ref{fig2}, $\beta_1$= 0.56 and $\beta_2= 2.23$ are obtained from Eq. (\ref{eq12}). Then with the help of Eq. (\ref{eq15}), the energy of the corresponding ground states are calculated as $E_1|_{g'=0.1}=-0.418$  corresponding to $\alpha_1$ and $\beta_1$ and  $E_2|_{g'=0.1}=0.463$  corresponding to $\alpha_2$ and $\beta_2$ i.e, there is one positive energy solution and one negative energy solution. We proceed to analyze the energy profile at ($\alpha_1$,$\beta_1$) and ($\alpha_2$,$\beta_2$) in the parameter space of $\alpha$ and $\beta$. We find that at ($\alpha_1$,$\beta_1$):
\begin{eqnarray}\label{eq16}
\begin{aligned}
\frac{\partial^2 E}{\partial\beta^2}=10.61;{~~}
\frac{\partial^2 E}{\partial\alpha^2}=1.66;{~~}
\frac{\partial^2 E}{\partial\alpha\partial\beta}=1.1
\end{aligned}
\end{eqnarray}
and at ($\alpha_2$,$\beta_2$)
\begin{eqnarray}\label{eq17}
\begin{aligned}
\frac{\partial^2 E}{\partial\beta^2}=0.84;{~~}
\frac{\partial^2 E}{\partial\alpha^2}=-437.75;{~~}
\frac{\partial^2 E}{\partial\alpha\partial\beta}=-0.22
\end{aligned}
\end{eqnarray}
Eq. (\ref{eq16}) and Eq. (\ref{eq17}) suggest that ($\alpha_1$,$\beta_1$) is the point of stable minima and ($\alpha_2$,$\beta_2$) is a saddle point in parameter space. This leads to the conclusion that the negative energy solution corresponding to ($\alpha_1$,$\beta_1$) is the only bound state energy. In Fig. \ref{fig3} we have plotted the energy profile in $\alpha$,$\beta$ parameter space.
\begin{figure}[!htbp]
\begin{center}
\includegraphics[angle=0,scale=0.8]{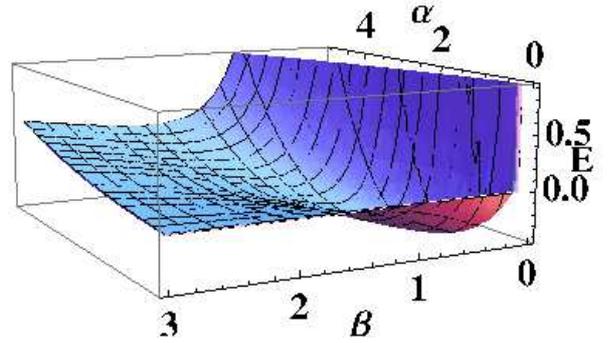}
\caption{In this figure we have plotted the energy profile ($E$ in the Z-axis is in the scale of $ND$) given in Eq. (\ref{eq15}) for $k=3$ and $g'=0.1$ in the $\alpha$,$\beta$ parameter space. This clearly signifies that ($\alpha_1$,$\beta_1$) is a global minima.}
\label{fig3}
\end{center}
\end{figure}
\begin{figure}
\includegraphics[angle=0,scale=0.5]{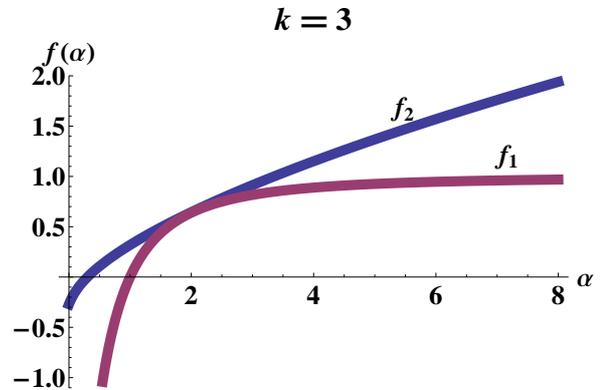}
\caption{This figure shows the plot of $f_1(\alpha)$ and $f_2(\alpha)$ vs. $\alpha$ for $k=3.0$ and $g'=0.17 =g'_c $. The single point of intersection at $\alpha= 2$ indicates the merge of two sets of solutions at $g'_c$ and appearence of 1 set of solution only for $\alpha$ and $\beta$. }
\label{fig4}
\end{figure}
The solution of $\alpha$ corresponds to the ground state energy $E|_{g'=g_c}=-0.31$ where $E|_{g_c}<E_1|_{g_{c-}}$ considering the magnitude only. This clearly signifies that with the increase of the value of $g'$ the localized nature of the system diminishes and the free particle character comes into play. 

\begin{figure}
\includegraphics[angle=0,scale=0.5]{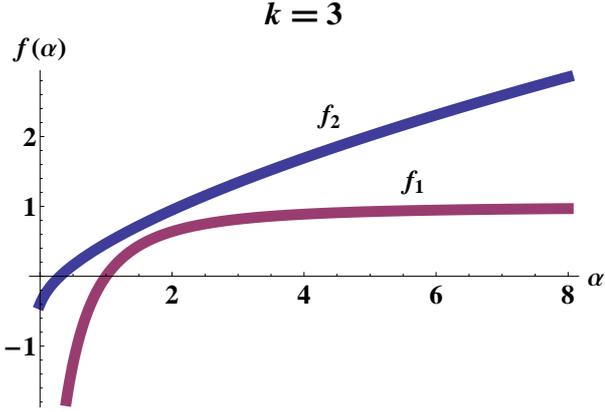}
\caption{Above figure shows the plot of $f_1(\alpha)$ and $f_2(\alpha)$ vs. $\alpha$ for $k=3.0$ and $g'= 1.0$. It shows that we have no solution for $\alpha$ at $g'>g'_c$. }
\label{fig5}
\end{figure}

If we increase $g'$ keeping $k$ fixed at 3.0, then after a certain value of $g'=g'_{c-}$, both $E_1$ and $E_2$ become negative. Further increase of $g'$, decreases the magnitude of $E_1$ and increases that of $E_2$ as illustrated in Table 1 and finally when $g'=g_c$, $f_2(\alpha)$ becomes tangent to $f_1(\alpha)$ as depicted in Fig. \ref{fig4}. Further increasing $g'$, there will be no solution for $\alpha>0$ (only $\alpha>0$ is physically relevant) which will be clear from Fig. \ref{fig5}. \\
\begin{table}[!htbp]
 \centering 
\begin{tabular}{c  c  c  } 
\hline\hline 
$g'$ & $E_1$ & $E_2$ \\ [0.5ex] 
\hline 
0.10 & -0.418 & 0.463\\
0.12 & -0.407 & 0.168\\
0.14 & -0.395 & -0.029\\
0.155 & -0.37 & -0.177\\ 
0.17\text{$=g'_c$} & -0.31 & -0.31\\[1ex]
\hline 
\end{tabular}
\caption{Results showing how the energy values changes with $g'$. $k$ is keeping fixed at 3.0. After $g'=0.12=g'_{c-}$ both $E_1$ and $E_2$ becomes negative and at $g'=g'_c=0.17$ only one bound state exist, $f_1$ and $f_2$ being tangent to each other and at $g'>g'_c$ no bound state exists at all.}
\end{table}
 
\subsection{Different values of $k$ is considered}

So far we have explained the case keeping $k$ fixed at 3.0 and now we will consider different values of $k$. The qualitative nature of the solutions remain the same as $k=3.0$ but the values of $g_c'$ and the ground state energy at $E|_{g_c'}$ changes with $k$ as is obvious from Table 2.  
\begin{table}[!htbp]
 \centering 
\begin{tabular}{c  c  c  } 
\hline\hline 
$k$ & $g'_c$  & $E_{g'_c}$  \\ [0.5ex] 
\hline 
2 & 0.445 & -0.048\\
3 & 0.170 & -0.310\\
4 & 0.095 & -0.459\\
5 & 0.061 & -0.546\\ [1ex]
 
%[1ex] adds vertical space
\hline 
\end{tabular}
\caption{Results showing how the critical value of coupling constants and the energies are varying for different values of $k$.}
\end{table}

From table 2, it can readily be said that with the increase of the value of $k$, critical value of $g$ decreases but $|E_{g_c'}|$ increases. $g_c'$, $|E_{g_c'}|$ approaches to the following limit: $g_c'\rightarrow 0$ and $E_{g_c'}\rightarrow-1$ as $k\rightarrow\infty$. Hence free particle nature of the bosons predominates more rapidly. The decrease of the values of ${g_c'}$ with the increase of $k$ implies that very small repulsive interatomic interactions at this point is sufficient to destroy the localized state. \\

From Eq. (\ref{eq8}) and Eq. (\ref{eq9}), it can readily be shown that for $k=1$, $\alpha$ is equal to 0 when there is no interaction present in the system. But we want the bound state of the form of Eq. (\ref{eq5}), hence our treatment will only be valid for $k\ge2$.
\begin{figure}[!htbp]
\includegraphics[angle=0,scale=0.6]{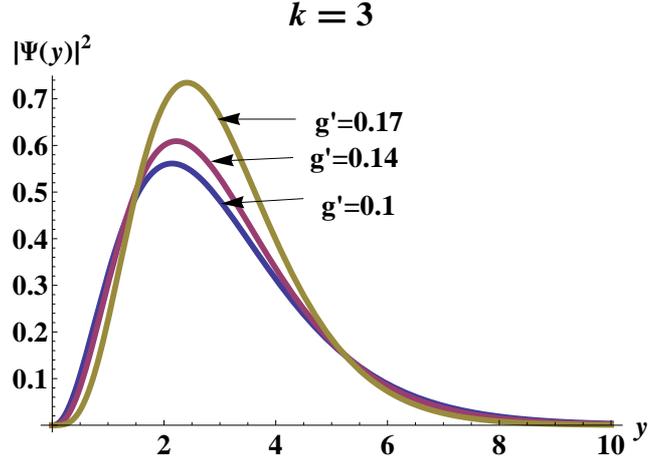}
\caption{Above figure shows the density profile of the localized state for different values of interaction strength $g'$  for $k=3$. In Y-axis $|\psi(y)|^2$ is in the scale and dimension of $\sqrt{\frac{NmD}{\hbar^2}}$. With the increase of $g'$, peaks are shifted towards right.}
\label{fig6}
\end{figure}
\begin{figure}[!htbp]
\includegraphics[angle=0,scale=0.6]{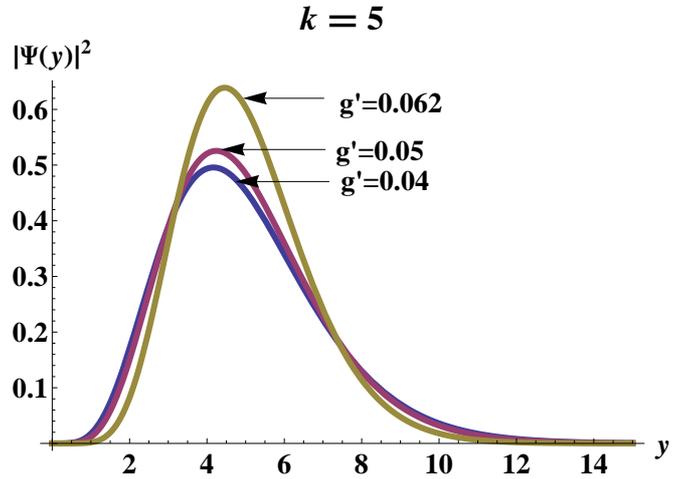}
\caption{This figure shows the density profile of the localized state for different values of $g'$ for $k=5$. The scale of Y axis remains the same as Fig. (\ref{fig6}).}
\label{fig7}
\end{figure}
\begin{figure}[!htbp]
\includegraphics[angle=0,scale=0.6]{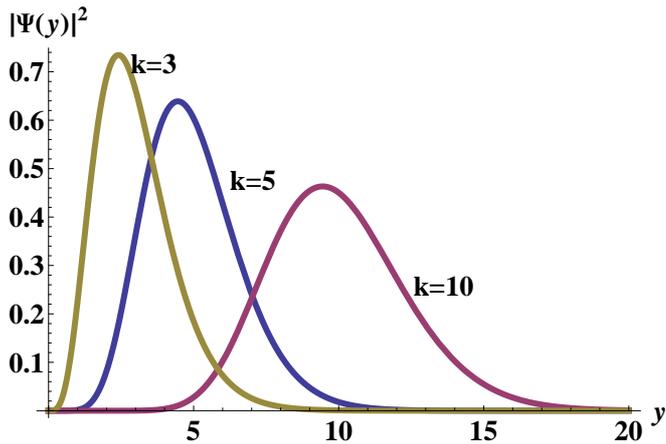}
\caption{This figure shows the density profile of the localized state at critical value of $g'$ i.e., at $g_c'$ for different values of $k$. The scale of Y axis remains the same as Fig. \ref{fig5}. The width of the curves increases with the increase of the value of $k$. }
\label{fig8}
\end{figure}

In Fig. \ref{fig6}, we have plotted $|\psi(y)|^2$ vs y for a particular value of $k$ $(k=3)$ which shows that with the decrease of the value of $g'$, the curve starts broadening which seems apparently counter-intuitive . But the careful observation reveals that though the peaks of the curves are around $y=2$, the peaks are going far apart from the trap centre as $g'$ is increased indicating higher repulsive interaction between the particles causing poor localization inside the trap and the particles are more prone to get rid of the confining potential. 

In Fig. \ref{fig7} also we have plotted the same for $k=5$ and we observe that the peaks are now located around $y=4$ and herealso the peaks are going far away from the trap centre with the increase of $g'$. Smaller magnitudes of $g'$ in this case than those of $k=3$ indicates that the small repulsive interaction are now sufficient to cause delocalisation. 

In Fig. \ref{fig8} we have shown the density profile at different values of $g_c'$ corresponding to different $k$'s. With the increase of the value of $k$, the increasing width of the curves indicates that as the system falls under weak confinement, the localised nature of the system at $g_c'$ diminishes.
\section{Conclusion}

As a summary, in this work starting from Gross-Pitaevskii energy functional we have explored analytically the ground state behavior of GPE which develops an interesting feature when the system is trapped under Morse potential. As the potential approaches to simple harmonic one, the system being highly trapped, the ground state solution becomes more and more localized, whereas in the opposite limit we have explained the free nature of the system dominates which is very relevant. This interesting limiting feature of the Morse potential is very unique. We have shown here for the first time that a bifurcation in the ground state solution appears when GPE is under the influence of Morse potential. Since GPE is nothing but NLSE, our work has indeed revealed an important characteristics of NLSE in Morse potential.\\

      This treatment can be extended to coupled Gross Pitaevskii equation also and in the limiting cases this should agree with the results which are already known in literature. 

\section*{Acknowledgements}
One of the authors, Sukla Pal would like to thank S. N. Bose National Centre for Basic Sciences for the financial support during the work. Sukla Pal acknowledges Harish-Chandra Research Institute for hospitality and kind support during visit.

\end{document}